\documentclass[11pt]{article}
\usepackage{moriond,epsfig}

\def\Journal#1#2#3#4{{#1} {\bf #2}, #3 (#4)}

\def\NPB{{\em Nucl. Phys.} B}

\def\PRL{\em Phys. Rev. Lett.}
\def\PRD{{\em Phys. Rev.} D}

\def\JPG{{\em Jour. Phys.} G}
\def\PRC{{\em Phys. Rev.} C}
\def\JPCS{\em Jour. Phys. Conf. Ser.}
\def\PR{\em Phys. Rep.}

\def\be{\begin{equation}}
\def\ee{\end{equation}}
\def\bea{\begin{eqnarray}}
\def\eea{\end{eqnarray}}

\def\raa{R_{AA}}
\def\piz{\pi^{0}}
\def\pt{p_{\mathrm{T}}}
\def\rg{R_{\gamma}}
\def\zt{z_{\mathrm{T}}}

\begin{document}
\vspace*{4cm}
\title{Aspects of Jet Production with PHENIX}

\author{ M. Nguyen for the PHENIX Collaboration}

\address{Department of Physics and Astronomy, Stony Brook University,\\
Stony Brook, NY, U.S.A.}

\maketitle\abstracts{
Measurement of the in-medium energy loss of fast partons is one of the most active topics in heavy-ion physics.  Such studies provide an opportunity to gain insight into the fundamental behavior of QCD processes by studying them away from vacuum conditions.  A promising channel for relating theoretical models to data are two particle correlations using direct photon triggers.  Recent results on this observeable using the PHENIX detector are presented.
}

\section{Parton Energy Loss in a QGP}
The primary motivation for nuclear collisions at the Relativistic
Heavy Ion Collider (RHIC) is to study the medium created at
high energy density, which we believe to be composed of a dense, thermalized,
effectively deconfined Quark-Gluon Plasma (QGP).  Jet tomography is
a promising tool with which to study medium properties.  In the
ideal case one would observe the attenuation of a parton beam of
fixed energy in a stationary sample of QGP, in analogy to X-ray
tomography in medical applications.  To the extent that the
interaction of fast partons with the medium is well understood, we
may then infer the density profile of the medium.  The theoretical
modeling of energy loss in hot nuclear matter, however, turns out to
be a rather rich field of study in itself.

In practice, we have access to neither parton beams nor QGP bricks.
In fact, the medium is small and fleeting, the size and lifetime
being only of order 10 fm.  Instead we use hard scattered partons
generated within the medium to probe the system.  The typical time
scale of jet production is such that partons are likely to lose
energy via gluon radiation as they traverse the medium and fragment
as normal, albeit at lower energy, outside the medium.  Then, at
least to lowest order, we might expect the observed distribution of
hadrons to reflect an effectively modified fragmentation function
which is simply shifted by the amount of lost energy which must be
averaged over all trajectories.


The calculation of $\Delta E$ for a given path-length, $L$, led to
the consideration of destructive interference of gluon
bremsstrahlung due to multiple scattering, the so-called
Landau-Pomeranchuk-Migdal (LPM) effect which was a previously
unsolved problem in QCD.  This effect predicts an $L^2$ dependence
to the $\Delta E$ as opposed to the linear dependence one naively
expects~\protect\cite{gyuwang}.  This provides a nice example of how we might test
fundamental predictions of QCD by introducing medium effects.

\begin{figure}
\begin{minipage}[t]{0.485\linewidth}
\centering
\includegraphics[width=8.5 cm]{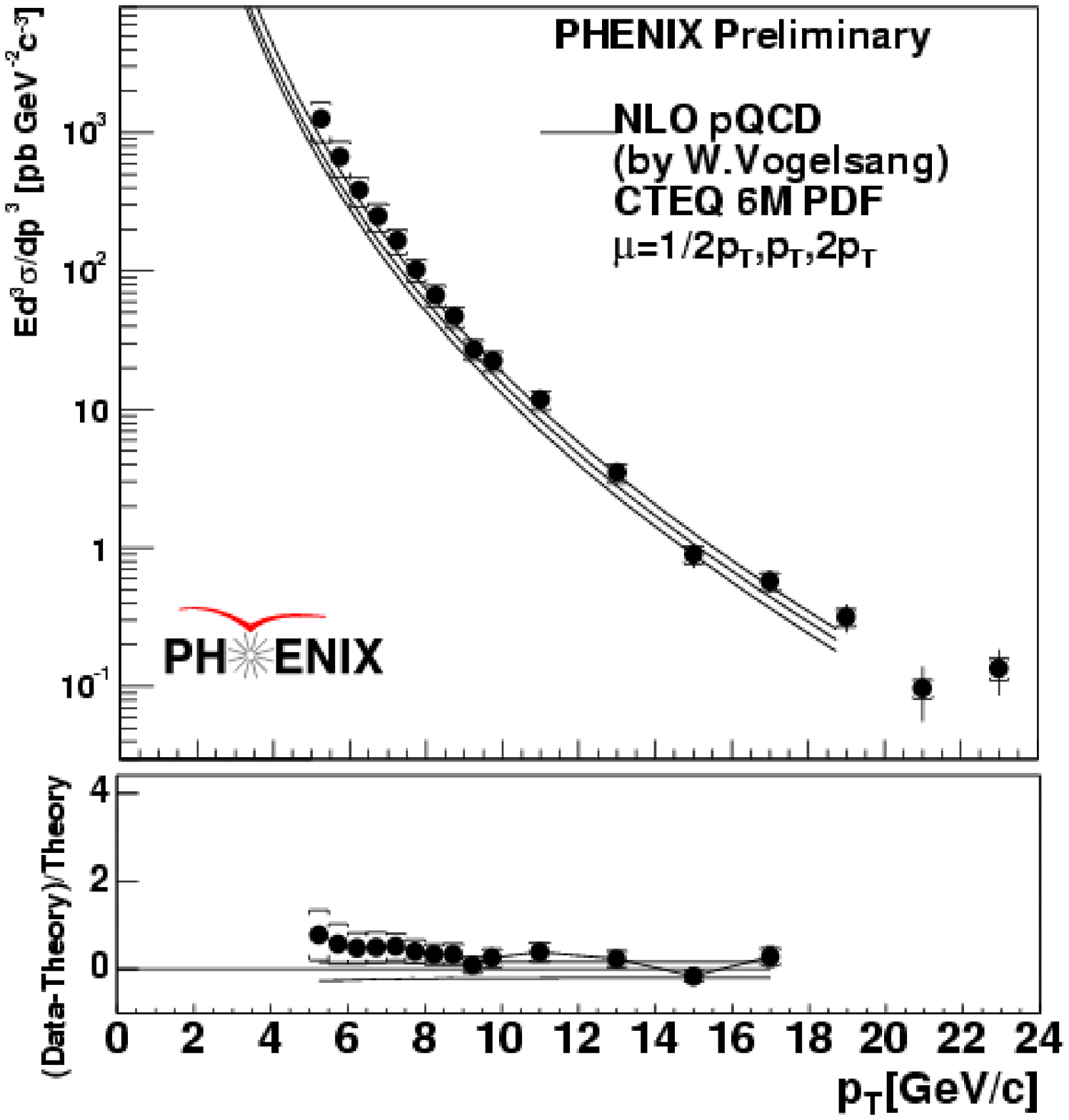} \vspace{0.02 mm}
\caption{\label{fig:run5pp_photon_sub}
Inclusive direct photon cross section in $p$+$p$ collisions compared to an NLO pQCD calculation~\protect\cite{wv}.
}
\end{minipage}
\hspace{0.3 cm}
\begin{minipage}[t]{0.485\linewidth}
\includegraphics[width=8.0 cm]{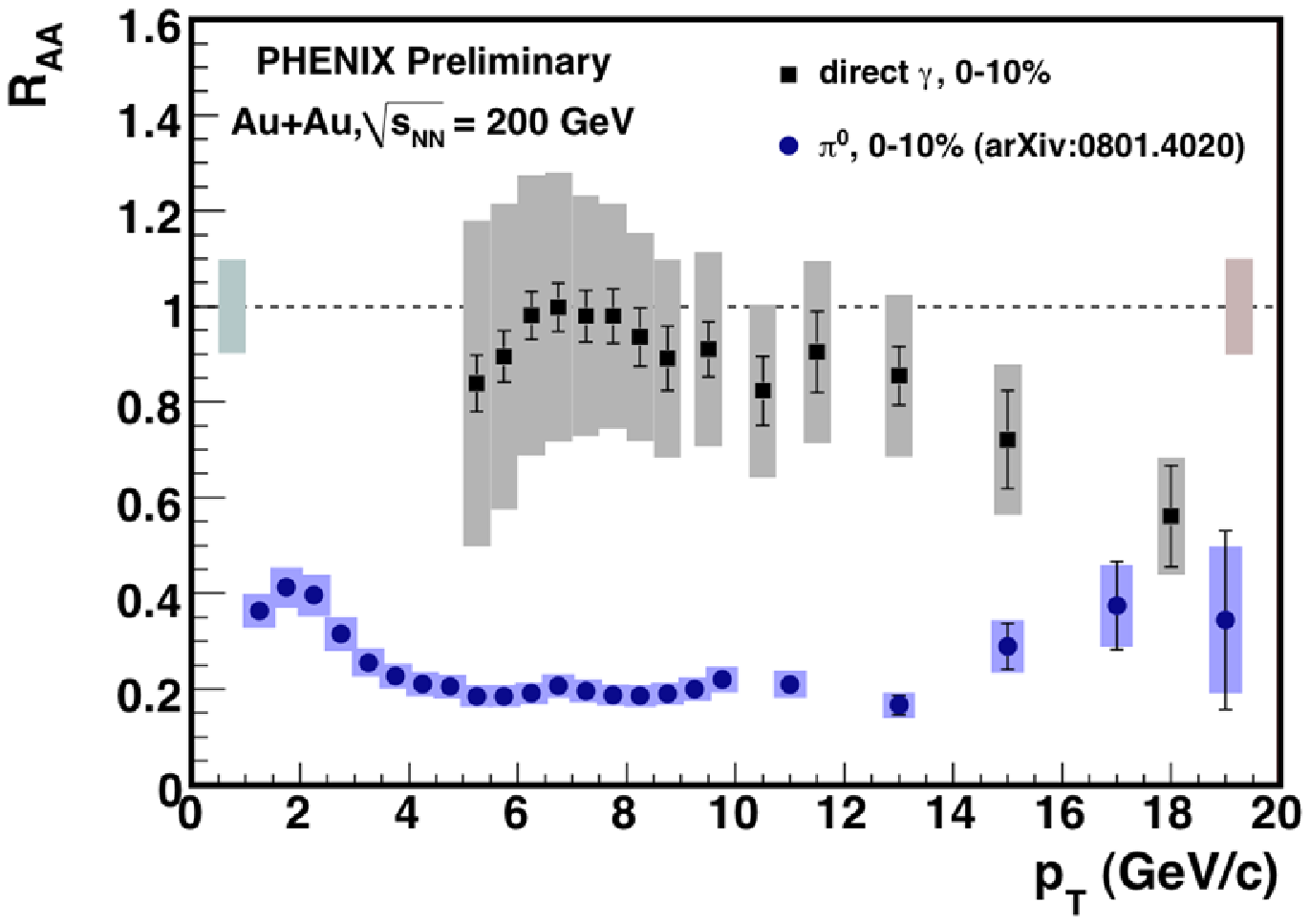} \hspace{3 cm}
\caption{\label{fig:RAA_pi0_dirpho}
Nuclear modification factor $\raa(\pt)$ for direct photons and $\piz$'s in central Au+Au collisions~\protect\cite{ppg080,reygers}.
}
\end{minipage}
\end{figure}

An accurate description of medium modification requires a thorough understanding of vacuum jet fragmentation which may be tested by comparing measured cross sections to Next-to-Leading Order (NLO) pQCD calculations.  Shown in Figure \ref{fig:run5pp_photon_sub} is the inclusive direct photon cross section in $p$+$p$ collisions which compare well with NLO calculations~\cite{wv}.

\section{High $\pt$ Suppression and Modified Jet Shapes}

Nuclear effects are quantified via the nuclear modification factor $\raa$ which is the ratio of the yield observed in Au+Au divided by the yield in $p$+$p$ collisions scaled by the number of binary collisions in Au+Au.  Shown in Figure \ref{fig:RAA_pi0_dirpho} is the $\raa$ for $\piz$ which shows a factor of five suppression which is taken as evidence that jets are strongly quenched by the medium~\protect\cite{ppg080}.  On the other hand, direct photons are unmodified as one would expect from a color neutral object which would not interact strongly with the medium~\protect\cite{reygers} \footnote{Note that at very high $\pt$ the data do show a hint of deviation from an $\raa$ of unity.  Such effects may be expected due to both initial and final state effects are reviewed in~\cite{gabor}.}.  

The best way to study jets is by employing jet reconstruction algorithms, a well developed procedure in the context of elementary particle collisions.  Such studies are indeed being pursued in heavy-ion collisions, but are complicated by the presence of a very large background.  Instead we focus on two particle correlations in which essential features of jet production are evident in azimuthal correlations between particle pairs.  Typically, we tabulate per-trigger yield ($Y$) of associated particles as a function $\Delta \phi$ from which a di-jet structure is manifest as a double peak structure.  The underlying event (UE) in $p$+$p$ collisions is treated as a pedestal and is removed by a two Gaussian + constant fit.  Although the true structure of the UE is known to be more complicated \cite{ue}, such a procedure is justified by the fact that a similar methodology is applied in Au+Au collisions and we are primarily interested in the difference between the two systems.  In Au+Au the UE is much larger due to the large number of soft collisions.  This background is subtracted by event mixing since, to good approximation, it is independent of the hard scattering.  The soft background does however contain correlations of its own due to the collective behavior of the system which must be subtracted away to study the jet correlations. 

In glancing (peripheral) Au+Au collisions, two particle correlations resemble the corresponding measurement in $p$+$p$ collisions.  In head-on (central) collisions, however, the away-side shape is drastically modified into a configuration in which the peak is shifted approximately one radian away from $\pi$, a feature colloquially referred to as the cone~\cite{ppg074}.  This observation spurred a tremendous amount of theoretical interest.  A number of theories have been proposed to explain the data.  For example, it's been argued that the away-side parton creates a mach cone as it passes through the medium~\cite{machcone}.

\begin{figure}
\begin{minipage}[tb]{0.485\linewidth}
\centering
\vspace{1 cm}
\includegraphics[width=8.0 cm]{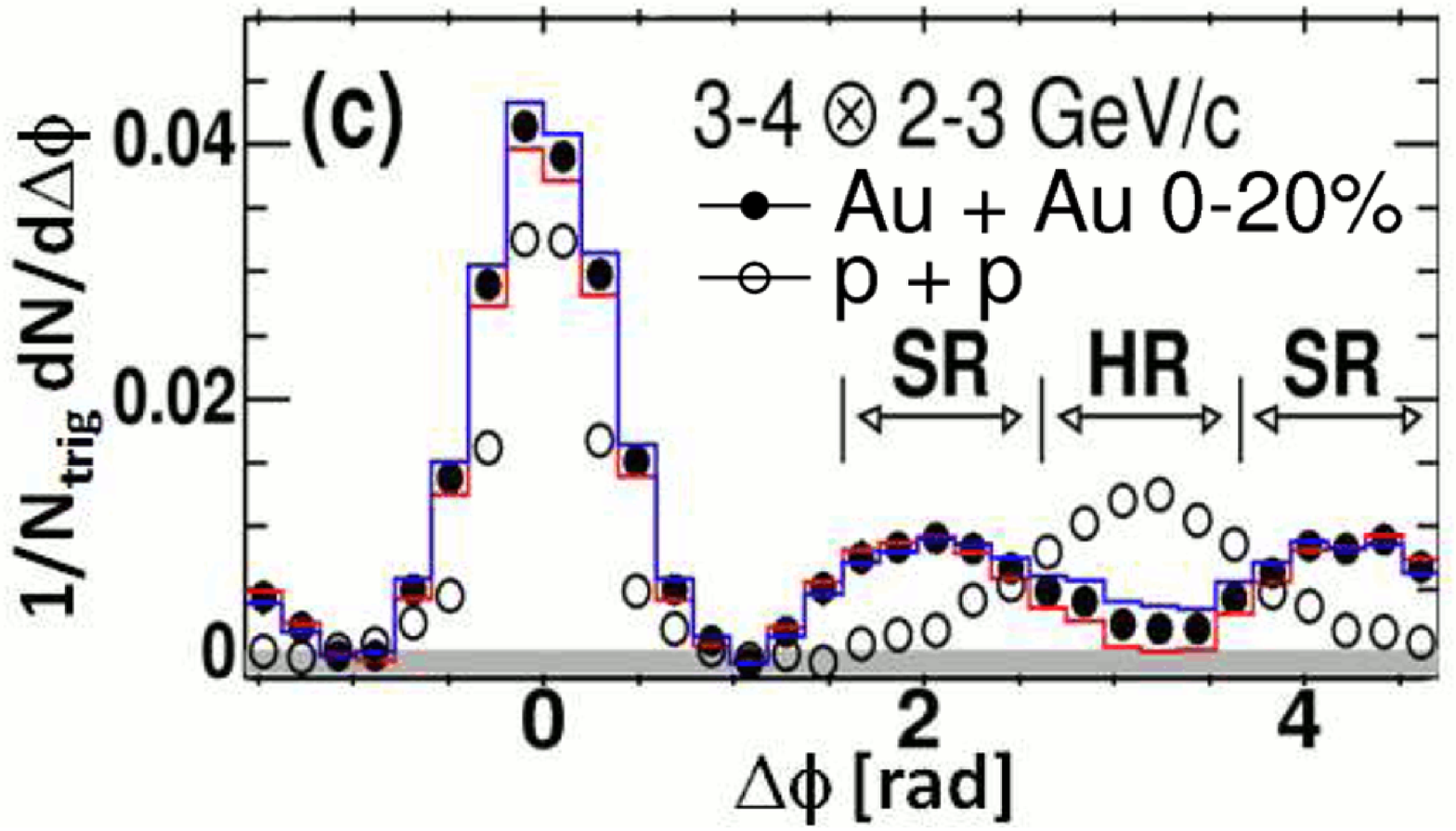} \vspace{0.1 cm}
\caption{\label{fig:cone}
Example of a di-hadron correlation measurement in which a modified jet shape is apparent~\protect\cite{ppg074}.
}

\end{minipage}
\hspace{0.3 cm}
\begin{minipage}[tb]{0.485\linewidth}
\centering
\includegraphics[width=8.5 cm]{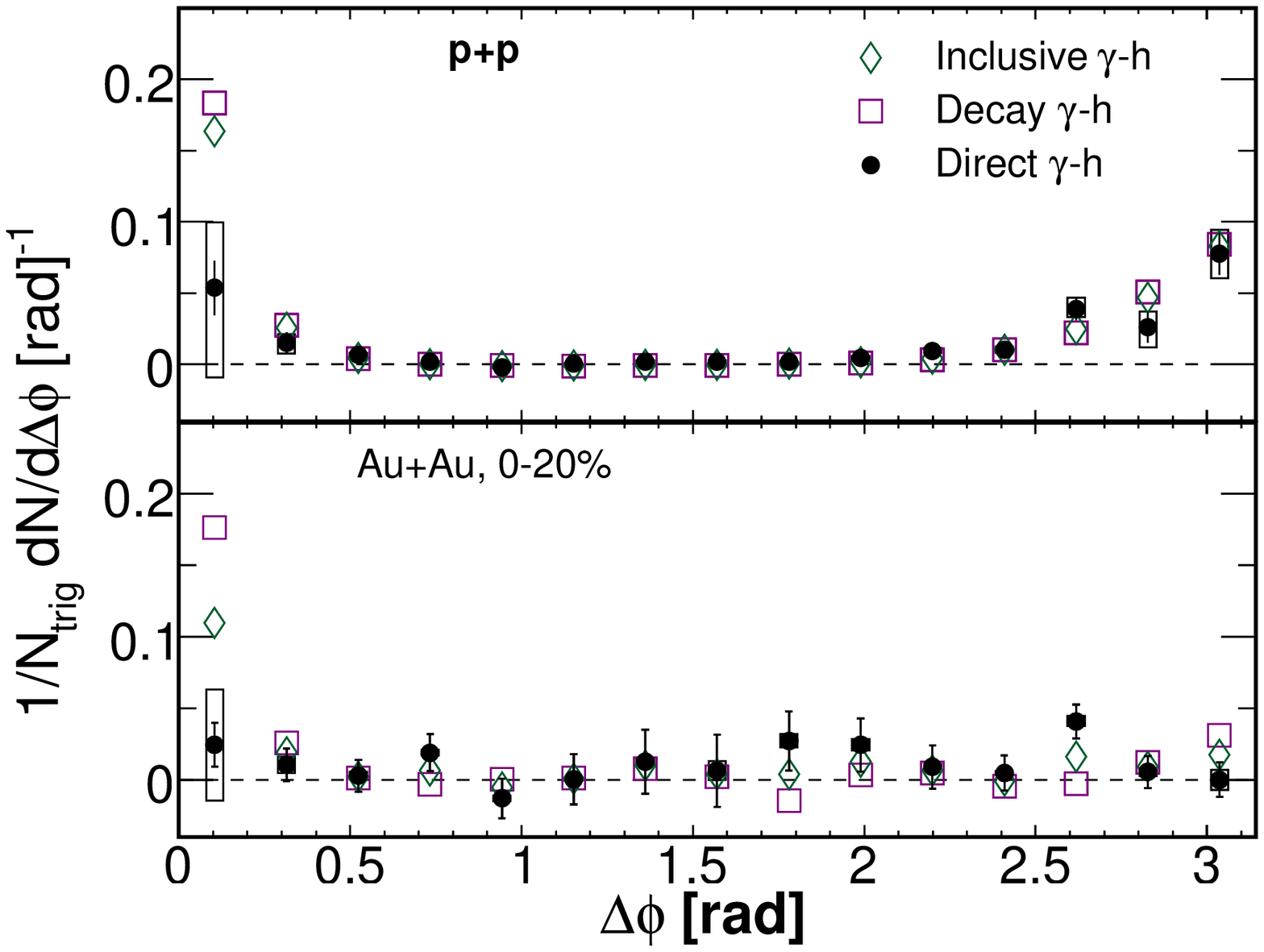}
\caption{\label{fig:dphi_incdecdir_ppg090}
Examples of direct photon triggered correlation measurements in $p$+$p$ and Au+Au collisions~\protect\cite{ppg090}.}
\end{minipage}
\end{figure}

An alternative explanation proposes that the mechanism of gluon radiation may itself be modified in such a way as to produce this feature.  This calculation uses standard perturbative methods, namely Sudakov form factors, to calculate in-medium jet shape modifications~\cite{polsal}.  However, an additional term is added to the parton splitting functions.  The authors find that large angle scattering is enhanced and, in principle, one is able to recover the cone-like jet shape.  Another approach, along similar lines, calculates not angular correlations, but rather the momentum distribution of final state hadrons using the so-called Modified Leading Log Approximation (MLLA) along with the assumption of Local Parton-Hadron Duality (LPHD)~\cite{borgwied}.  This calculation predicts a characteristic enhancement of particles at low fractional momentum which can be tested at RHIC and even better at the larger jet energies available at the LHC.  More advanced calculations are currently being implemented into Monte Carlo simulations 
however reliable predictions are for di-hadron correlations are difficult to obtain as the trigger bias introduced by such measurements is problematic.  For quantitative comparisons we must look to full jet reconstruction or direct photon correlations where the initial jet energy may be determined.

\section{Direct Photon Correlations}

Shown in Figure \ref{fig:dphi_incdecdir_ppg090} are examples of two particle correlations using direct photon triggers and associated charged hadrons in both $p$+$p$ and Au+Au.  The analysis is performed by measuring the per-trigger yield of inclusive photons and estimating that of decay photons from the measured $\piz$ and $\eta$ triggered correlations according to a Monte Carlo based calculation as described in~\cite{ppg090}.  The direct photon correlations are obtained by a statistical subtraction of these two quantities according to

\begin{equation}
Y_{direct} = \frac{\rg}{\rg-1} Y_{inclusive} - \frac{1}{\rg-1} Y_{decay},
\end{equation}

\noindent where $\rg$ ($\equiv N_{inclusive}/N_{decay}$) is determined from the decay and direct photon spectra.

\begin{figure}
\begin{minipage}[tb]{0.5\linewidth}
\centering
\includegraphics[width=8.5 cm]{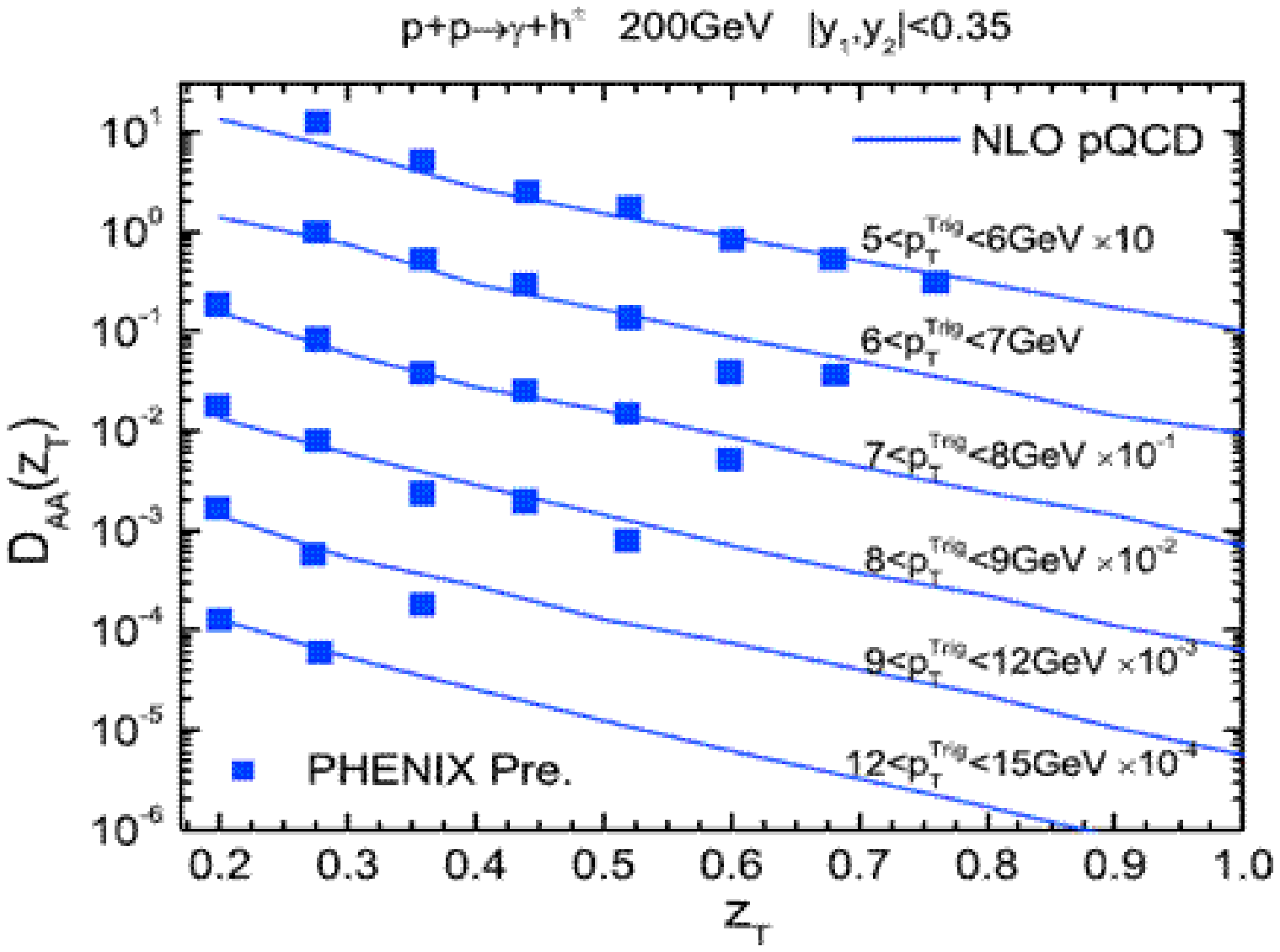}
\caption{\label{fig:NLO}
Away-side charged hadron yield per isolated direct photon and NLO calculations from~\protect\cite{zoww}.
}
\end{minipage}
\hspace{0.3 cm}
\begin{minipage}[tb]{0.475\linewidth}
\centering
\vspace{5mm}
\includegraphics[width=8.0 cm]{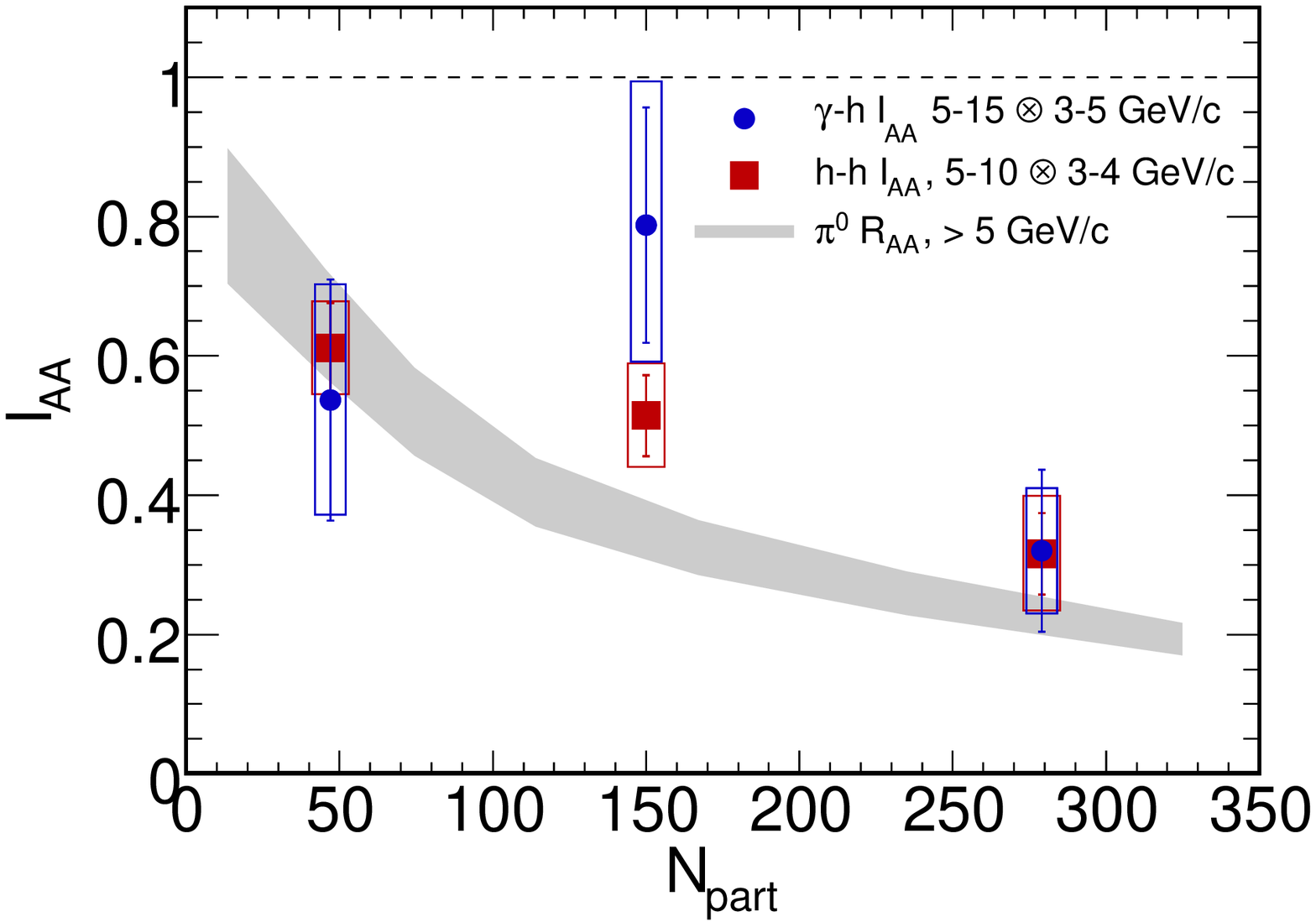}
\caption{\label{fig:integrals_vscent}
$\gamma$-$h$ $I_{AA}$~\protect\cite{ppg090}, $\piz R_{AA}$~\protect\cite{ppg080} and $h$-$h$ $I_{AA}$~\protect\cite{ppg083} \emph{vs.} number of participants for the $\pt$ selections as indicated. 
}
\end{minipage}
\end{figure}

One can test the compatibility of the measurement with NLO calculations.  Shown in Figure \ref{fig:NLO} is the $\zt$ ($\equiv p_{T}^{h}/p_{T}^{\gamma}$) distribution for isolated direct photons~\protect\cite{frantzHP} in $p$+$p$ collisions where, modulo the $k_{\mathrm{T}}$ effect, the distribution measures the the fragmentation function of the away-side jet.  Also shown are NLO calculations from \cite{zoww} using the KKP parametrization \cite{kkp} of the FF's which show good agreement with the data.


  Figure \ref{fig:integrals_vscent} shows $\gamma$-$h$ $I_{AA}$ for the ratio of the per-trigger yield in Au+Au to that in $p$+$p$ collisions, for the away-side head region (HR) as shown in Figure \ref{fig:cone}.  The $\pt$ selection, as indicated in the legend, corresponds to $\langle \zt \rangle \approx 0.45$.  The similarity to the large suppression shown by the high $\pt$ single $\piz$ yield ($R_{AA}$) suggests that surface emission is dominant and supports a picture in which the medium is composed of an extremely opaque core. Depending on the density profile, one might expect di-hadron correlations to show a different level of suppression since the both jets are biased towards small energy loss.  Within the statistical precision of the measurement, however, no difference is apparent.

 Model calculations suggest that sensitivity to the energy loss mechanism and density profile are maximized at low values of $\zt$ where the $\langle \Delta E \rangle$ of the away-side jet becomes large~\protect\cite{renkgjet,borgwied,zoww}.   New data from recent and upcoming runs should provide the improved statistical precision to measure to low values of $\zt$ where and to confront the energy loss models discussed in this work.

\end{document}